\documentclass[preprint, aps, prd, superscriptaddress, nofootinbib, floatfix]{revtex4-2}

\usepackage[T1]{fontenc}
\usepackage[utf8]{inputenc}
\usepackage{amsmath, amssymb, physics, bm}
\usepackage{graphicx, xcolor}
\usepackage{siunitx}
\usepackage{ulem}
\usepackage{booktabs}
\usepackage{comment}
\usepackage{stackengine}
\usepackage{adjustbox}
\usepackage{slashed}
\usepackage{multirow}
\usepackage{caption}
\usepackage{geometry,tabularx}
\usepackage{fontawesome} 

\usepackage{subcaption}

\usepackage[dvipsnames]{xcolor}

\usepackage[colorlinks=true, linkcolor=blue, citecolor=blue, urlcolor=blue]{hyperref}
\usepackage{cleveref}

\begin{document}

\flushbottom 

\title{\Large Temperature-Dependent CPT Violation: Constraints from Big Bang Nucleosynthesis}

\author{Gabriela Barenboim}
\affiliation{Instituto de F\'{i}sica Corpuscular, CSIC-Universitat de Val\`{e}ncia, Paterna 46980, Spain}
\affiliation{Departament de F\'{i}sica Te\`{o}rica, Universitat de Val\`{e}ncia, Burjassot 46100, Spain}

\author{Anne-Katherine Burns}
\affiliation{Departament de Física Quàntica i Astrofísica (FQA), Universitat de Barcelona (UB), c. Martí i Franqués, 1, 08028 Barcelona, Spain}
\affiliation{Institut de Ciències del Cosmos (ICCUB), Universitat de Barcelona (UB), c. Martí i Franqués, 1, 08028 Barcelona, Spain}

\email{gabriela.barenboim@uv.es}
\email{annekatherineburns@icc.ub.edu}

\begin{abstract}
In this study, we explore temperature-dependent CPT violation during Big Bang Nucleosynthesis (BBN) through electron-positron mass asymmetries parametrized by $b_0(T) = \alpha T^2$. The $T^2$ scaling naturally evades stringent laboratory bounds at zero temperature while allowing for significant CPT violation at MeV scales in the early universe \cite{ParticleDataGroup:2024cfk}. Using a modified version of the BBN code \faGithub \href{https://github.com/vallima/PRyMordial}{\,\texttt{PRyMordial}} with dynamically-solved chemical potentials and appropriate finite-mass corrections, we constrain electron-positron mass differences from observed abundances of Helium-4, Deuterium, and $N_{\rm eff}$. We find that $\alpha$ must be greater than or approximately equal to $10^{-6}$ GeV$^{-1}$ for keV-scale mass differences at BBN. All three observables show no simultaneous $1\sigma$ overlap, though pairwise combinations allow for constrained regions of parameter space. We present two toy models demonstrating how $b_0(T) \propto T^2$ arises from field-theoretic mechanisms, including temperature-driven phase transitions. These results provide the most stringent constraints on early-universe CPT violation in this regime, probing parameter space inaccessible to laboratory experiments.
\end{abstract}

\maketitle

\pagebreak

\section{Introduction}

Charge-Parity-Time (CPT) symmetry is a fundamental cornerstone of quantum field theory, requiring that particles and antiparticles have identical masses and lifetimes. Experimental tests have verified CPT invariance to remarkable precision at low energies, with constraints on electron–positron mass differences reaching extraordinary levels in present-day 
laboratory experiments \cite{Gabrielse:2025jep,CMS:2021jnp,STAR:2019wjm,KLOE-2:2018yif,Barenboim:2017ew,RavonelSalzgeber:2015wna,ALICE:2015rey,ParticleDataGroup:2024cfk}. However, these stringent bounds apply only to the present-day universe and do not necessarily constrain CPT violation in the early universe, during which time temperatures were many orders of magnitude higher.

A key open question is whether CPT symmetry could have been violated at high temperatures in the early universe while remaining unobservable today. In particular, a temperature-dependent mass difference between electrons and positrons caused by CPT violation during Big Bang Nucleosynthesis (BBN) would modify weak interaction rates, thus altering the ratio of the number of neutrons to the number of protons at freeze out, and in addition, change the evolution of the temperature of the thermal bath, both of which lead to a shift in the resulting primordial element abundances. In 2005, Lambiasea studied the effect of CPT and Lorentz invariance breakdown on the primordial Helium-4 abundance, using the neutron to proton ratio to estimate the Helium-4 mass fraction \cite{Lambiase:2005kb}. However, early-universe constraints on such mass asymmetries have not yet been fully mapped out in a framework that incorporates realistic thermal evolution. 

Most existing analyses assume constant mass differences or treat CPT violation as a temperature-independent low-energy parameter \cite{Dolgov:2014xza, Barenboim:2017vlc, Dolgov:2009yk, Bolokhov:2006wu, Bossingham:2017gtm}. These approaches cannot capture the possibility that mass asymmetries arise dynamically through finite-temperature effects or evolving background fields. Without accounting for temperature dependence, current constraints incorrectly estimate the true viability of CPT-violating scenarios during BBN.

In this work, we explore the possibility of temperature-dependent CPT violation that manifests as a mass difference between electrons and positrons during BBN. Such a scenario could have significant implications for the evolution of the primordial element abundances as well as the predicted effective number of neutrino species, $N_{\rm eff}$, at the end of BBN. We use the precision BBN code \faGithub \href{https://github.com/vallima/PRyMordial}{\,\texttt{PRyMordial}} to systematically map out constraints on electron and positron masses from observed abundances of Helium-4, Deuterium, Lithium-7, and measurements of $N_{\rm eff}$ \cite{Burns:2023sgx}. 

We introduce a framework in which the electron–positron mass difference is controlled by a temperature-dependent background parameter $b_0(T)$, motivated by finite-temperature field theory and high-scale symmetry breaking. In particular, we consider scalings such as $b_0(T) \propto T^2$, which naturally allow sizable early-universe CPT violation while ensuring that present-day experimental bounds are automatically satisfied. We note that here, $b_0(T)$  denotes a generic CPT-violating background parameter and should not be identified with the SME axial coefficient $b_\mu^{SME}$.

The temperature-dependent form of the potential is crucial, because it captures the generic behavior expected from finite-temperature field theory: thermal backgrounds modify dispersion relations, shift effective masses, and can induce temporary CPT-violating terms that vanish as $T \to 0$. In this sense, a temperature-dependent $b_0(T)$ is not an ad hoc choice but a minimal assumption consistent with thermal corrections in a relativistic plasma. 

The parameter $b_0(T)$ acts as a CPT-violating background field that shifts the dispersion relations of electrons and positrons asymmetrically. Its value controls the induced mass splitting and may naturally evolve with temperature due to thermal expectation values or symmetry-breaking dynamics.

The specific scaling $b_0(T) \propto T^2$ represents the maximum temperature dependence compatible with both theoretical consistency and phenomenological viability. Alternative scalings with higher powers---such as $b_0(T) \propto T^3$ or $T^4$---would grow too rapidly at high temperatures, producing enormous CPT violation at BBN energies ($T \sim 1$ MeV) that would grossly violate observational constraints even for small coupling constants. To illustrate: for $T^4$ scaling with $b_0(T) = \alpha T^4$ and $\alpha \sim 10^{-16}~\text{GeV}^{-3}$, one would obtain $b_0(T_{\rm BBN}) \sim 10^{-4}$ GeV at BBN temperatures, corresponding to $\sim 100$ keV mass differences---far exceeding any phenomenologically viable range. $T^3$ scaling would be similarly problematic, though requiring somewhat less extreme values of the coupling constant.

Conversely, weaker scalings such as $b_0(T) \propto T$ are either incompatible with standard finite-temperature power counting (for higher-dimensional operators, the leading thermal contribution naturally scales as $T^2$) or produce effects too small to be phenomenologically interesting at BBN, while still conflicting with present-day laboratory bounds. The $T^2$ behavior thus occupies a unique window: it is (i) the steepest scaling that remains phenomenologically viable at BBN, (ii)  emerges naturally from finite-temperature effective potentials and thermal loop corrections, and (iii) sufficiently strong to vanish rapidly enough as $T \to 0$ to satisfy present-day CPT tests. This makes $T^2$ both the theoretically minimal and phenomenologically optimal choice.

In a temperature-dependent scenario, CPT violation can be significant at $T \sim \text{MeV}$ while rapidly diminishing as the universe cools, thereby avoiding conflict with laboratory constraints. This resolves the tension faced by constant-mass-difference models and provides a physically motivated mechanism for early-universe CPT violation. By connecting this framework to precise BBN predictions, we identify the regions of temperature-dependent parameter space that remain phenomenologically viable.

The purpose of this paper is to use BBN observations to constrain the parameter $\alpha$ in the relation $b_0(T) = \alpha T^2$. For large values $\alpha \gtrsim 10^{-6}~\text{GeV}^{-1}$, mass differences become significant (keV-scale or larger) at BBN temperatures, leading to observable modifications of primordial abundances and $N_{\rm eff}$ that can be tightly constrained by precision cosmology, as explored in Sections~\ref{sec:BBN_constraints} and~\ref{sec:theory_models}. These constraints represent the most stringent bounds on early-universe CPT violation of this type. In addition, we note that the same theoretical framework with much smaller coupling constants $\alpha \sim 10^{-10}~\text{GeV}^{-1}$ could in principle address baryogenesis through sphaleron conversion at the electroweak scale, though this scenario is physically independent from and incompatible with the large-$\alpha$ regime constrained by BBN.

\section{Thermal Equilibrium with Mass Asymmetry}

\subsection{Charge Neutrality and Chemical Potential}

In a thermodynamic system containing electrons and positrons, thermal equilibrium implies that both species possess well-defined energy distributions governed by Fermi-Dirac statistics. The occupation number for a fermionic state of energy $E$ is given by
\begin{equation}
    f(E) = \frac{1}{e^{(E - \mu)/T} + 1},
\end{equation}
where $T$ is the temperature, $\mu$ is the chemical potential associated with the particle species, and the energy is related to the momentum via the relativistic dispersion relation $E = \sqrt{p^2 + m^2}$.

Under standard CPT-invariant assumptions, the electron and positron are exact antiparticles: they have equal masses, opposite electric charges, and chemical potentials that satisfy $\mu_{e^+} = -\mu_{e^-}$. In such a system, electric charge neutrality requires that the number densities of electrons and positrons be equal, $n_{e^-} = n_{e^+}$, since their charges are equal in magnitude and opposite in sign.

If CPT symmetry is violated, and the electron and positron acquire different masses: $m_e$ for the electron and $m_{e^+} \ne m_e$ for the positron. This asymmetry modifies the dispersion relation for each species but does not affect the fundamental form of the thermal distribution. Each species continues to follow the Fermi-Dirac distribution:
\begin{align}
    f_{e^-}(p) &= \frac{1}{e^{(\sqrt{p^2 + m_e^2} - \mu)/T} + 1}, \\
    f_{e^+}(p) &= \frac{1}{e^{(\sqrt{p^2 + m_{e^+}^2} + \mu)/T} + 1},
\end{align}
in which we have used the condition for chemical equilibrium that arises from the reaction $e^- + e^+ \leftrightarrow 2\gamma$, which imposes
\begin{equation}
    \mu_{e^-} + \mu_{e^+} = 0.
\end{equation}

The mass difference enters here only through the energy-momentum relation in the distribution function. Each species maintains a thermal spectrum characterized by the common temperature $T$ and chemical potential structure $\mu_{e^-} = \mu$, $\mu_{e^+} = -\mu$.

Since U(1) electromagnetic gauge symmetry is preserved, charge neutrality is automatically maintained in the system. The number density of each species is given by
\begin{equation}
    n = g \int \frac{d^3p}{(2\pi)^3} \frac{1}{e^{(E - \mu)/T} + 1},
\end{equation}
where $g = 2$ accounts for the spin degeneracy of the electron and positron. If $m_e \ne m_{e^+}$, the number densities of $e^-$ and $e^+$ are no longer equal for arbitrary values of $\mu$. The chemical potential is fixed dynamically by the charge neutrality condition:
\begin{equation}
    n_{e^-}(m_{e}, \mu, T) = n_{e^+}(m_{e^+}, -\mu, T),
\end{equation}
in which we neglect the contribution from tiny net positive charge of baryons and nuclei. This equation determines the value of $\mu$ that the system naturally adopts. In effect, the chemical potential compensates for the mass asymmetry between the species, ensuring charge neutrality is preserved without any additional constraint.

\subsection{Limiting Behaviors}

In the relativistic limit, where $T \gg m_e, m_{e^+}$, the effect of the mass difference becomes negligible, and the chemical potential required to ensure charge neutrality tends toward zero. In contrast, in the non-relativistic regime $T \ll m$, the number densities acquire exponential suppression:
\begin{align}
    n_{e^-} &\propto e^{-(m_e - \mu)/T}, \\
    n_{e^+} &\propto e^{-(m_{e^+} + \mu)/T}.
\end{align}
Charge neutrality in this limit for a pure $e^{\pm}$ plasma with no other charged species, requires that the chemical potential satisfy
\begin{equation}
    m_e - \mu = m_{e^+} + \mu \quad \Rightarrow \quad \mu = \frac{1}{2}(m_e - m_{e^+}).
\end{equation}
Thus, even in the non-relativistic case, charge neutrality is preserved by a chemical potential shift proportional to the mass difference.

An approximation valid in the temperature range of interest, T $\sim$ 0.01 - 10 MeV, is:

\begin{equation}
\mu(T) \approx \frac{T}{2} \ln \left( \frac{m_{e^+}^2 \, K_2\left(\frac{m_{e^+}}{T}\right)}{m_e^2 \, K_2\left(\frac{m_e}{T}\right)} \right),
\label{eq:mu_approx}
\end{equation}

where $K_2(z)$ is the modified Bessel function of the second kind, which naturally appears in thermal integrals for relativistic particles \cite{Giudice:2000ex}. This approximation correctly interpolates between the relativistic limit ($\mu \to 0$) and the non-relativistic limit ($\mu \to \frac{1}{2}(m_e - m_{e^+})$). In our analysis, in order to determine the most rigorous constraints on our model, we numerically solve for the chemical potential as a function of temperature. We have found that the resulting chemical potential is in good agreement with Equation \ref{eq:mu_approx}.

In conclusion, CPT violation leading to a mass asymmetry between electrons and positrons does not preclude thermal equilibrium, nor does it require a distortion of the thermal spectra. Both species retain their ideal Fermi-Dirac distributions, each shaped by their respective mass. Charge neutrality is maintained by adjusting the chemical potential so that the number densities of electrons and positrons remain equal, regardless of whether the particles are relativistic or non-relativistic. The chemical potential plays a central role in reconciling thermal equilibrium with charge neutrality in this CPT-violating context. This is a collective effect: the electron chemical potential arises to maintain conserved quantities, particularly electric charge neutrality, in thermal equilibrium. It is not a microscopic property of individual particles but a collective thermodynamic parameter describing the ensemble.

\section{Constraints from Big Bang Nucleosynthesis}
\label{sec:BBN_constraints}

\subsection{Methodology}

Through spectroscopic analysis of young, metal-poor galaxies we can determine the primordial abundances of Helium-4 and Deuterium \cite{ParticleDataGroup:2024cfk, Matsumoto:2022tlr}. Using this data, along with the most recently determined value of $N_{eff}$ we are able to constrain the parameter $\alpha$ in the relation $b_0(T) = \alpha T^2$ using the BBN code \faGithub \href{https://github.com/vallima/PRyMordial}{\,\texttt{PRyMordial}}. 

The following modifications have been made to \faGithub \href{https://github.com/vallima/PRyMordial}{\,\texttt{PRyMordial}} to encode the physics of dynamical CPT violation:

\begin{itemize}
    \item The electron and positron energy density and pressure, which are later used to calculate the thermodynamics of the background, are calculated independently.
    \item The electron/positron chemical potential is solved for dynamically such that charge neutrality in the early universe is maintained.
    \item The full calculation of the weak rates starting from the Born Approximation and including radiative corrections, finite mass effects, and thermal corrections is modified to allow for changes in the electron and positron masses \cite{Burns:2023sgx, Pitrou:2018cgg}.
    \item The interactions between electrons, positrons and neutrinos are modified appropriately. Specifically, the collision terms in the Boltzmann equation for energy densities of the neutrinos include corrections relating to the effect of finite electron mass in electron-neutrino scattering and annihilation matrix elements from Ref. \cite{Escudero:2025kej}. 
    \item Ref. \cite{Escudero:2025kej} provides QED plasma corrections which have been calculated using the standard electron and positron masses. These corrections play a role in the calculation of the total entropy, Hubble, and the plasma temperature evolution and are modified via a rescaling of the precomputed interaction pressure and its derivatives.
    \item The nuclear rates are normalized using the standard value of the electron mass, as opposed to the modified one as they are calculated using the measured values of atomic mass excess values for neutral atoms from NUBASE2020 \cite{Kondev_2021}.
\end{itemize}

\subsection{Observational Data}

The primordial abundances of both Helium-4 and Deuterium are measured via spectroscopy using emission and absorption lines from young, metal poor galaxies. Specifically, Deuterium is measured in absorption spectra by observing the spectra's isotope shifted Lyman-$\alpha$ features. BBN is the only known significant source of Deuterium in the universe and in addition, it is known to be entirely destroyed in stellar processing. Thus, measured values are lower bounds on the primordial abundance. The PDG's recommended value for the observed abundance of Deuterium is,

\begin{equation}
D / H \times 10^5 = (2.547 \pm 0.029) \quad \text{\cite{ParticleDataGroup:2024cfk}.}
\end{equation}

This value is the weighted average of 16 measurements of the Deuterium abundance and is in tension at the level of 2$\sigma$ with the Standard Model (SM) predicted value \cite{ParticleDataGroup:2024cfk}.

The primordial Helium-4 abundance is determined via the combined emission lines of Hydrogen and Helium in low-redshift extremely metal poor galaxies (EMPGs) and metal poor galaxies (MPGs). The measured value is extrapolated down to zero-metallically to get the true primordial value. The PDG recommended value of the primordial Helium-4 abundance is, 

\begin{equation}
Y_p = (0.245 \pm 0.003) \quad \text{\cite{ParticleDataGroup:2024cfk}.}
\end{equation}

In 2022 the EMPRESS collaboration published a new result for the primoridal Helium-4 abundance. Their analysis included data from 10 newly observed EMPGs with less than one-tenth the metal content of the sun. They combined this data with existing data from three EMPGs and 51 MPGs with metal content between one and four tenths that of the sun which was used to determine the PDG value above. Their resulting value is in tension with the SM prediction for $Y_p$ at the level of 3$\sigma$: 

\begin{equation}
Y_p = (0.2370^{+0.0034} _{-0.0033}) \quad \text{\cite{Matsumoto:2022tlr}.}
\end{equation}

The primordial abundance of Lithium-7 is measured in the atmospheres of metal-poor stars. The observed value has a significantly higher relative uncertainty as compared to the uncertainties on the observed value of the primoridal Helium-4 and Deuterium abundances, and is in tension with the SM predicted value at the level of 5$\sigma$. This discrepancy is widely known as the "Lithium Problem" to which many solutions have been proposed \cite{Fields:2011zzb, Miranda:2025wcp, Makki:2024sjq, Ali:2022moz, Koren:2022axd, Franchino-Vinas:2021nsf, Hayakawa:2021jxf, Deal:2021kjs, Hou:2021ynb, Anchordoqui:2020djl, Clara:2020efx, Flambaum:2018ohm, Hou:2017uap, Salvati:2016jng, Yamazaki:2014fja}. A recent and well motivated proposal has come from Brian Fields and Keith Olive in their 2022 paper \cite{Fields:2022mpw}. 

The recommended PDG value for the primordial abundance of Lithium-7 is:

\begin{equation}
^7Li / H \times 10^{10} = (1.6 \pm 0.3) \quad \text{\cite{ParticleDataGroup:2024cfk}.}
\end{equation}

The latest value of $N_{eff}$, the effective number of neutrino species comes from a combination of the ACT Data Release 6 power spectra, the Planck legacy CMB spectra, BAO from DESI Year-1 data, and CMB lensing from both Planck and ACT. This value is independent of BBN measurements and is determined by observing alteration in the damping tail of the TT/TE/EE spectra. The new value, assuming a fixed total neutrino mass of 0.06 eV, is:

\begin{equation}
N_{eff} = (2.86 \pm 0.13) \quad \text{\cite{ACT:2025tim}.}
\end{equation}

This value is in slight tension with the SM value of 3.044.

\subsection{Initial Results: Equal Mass Variation}

\begin{figure}[ht]
    \centering
    \begin{subfigure}{0.48\textwidth}
        \centering
        \includegraphics[width=\linewidth]{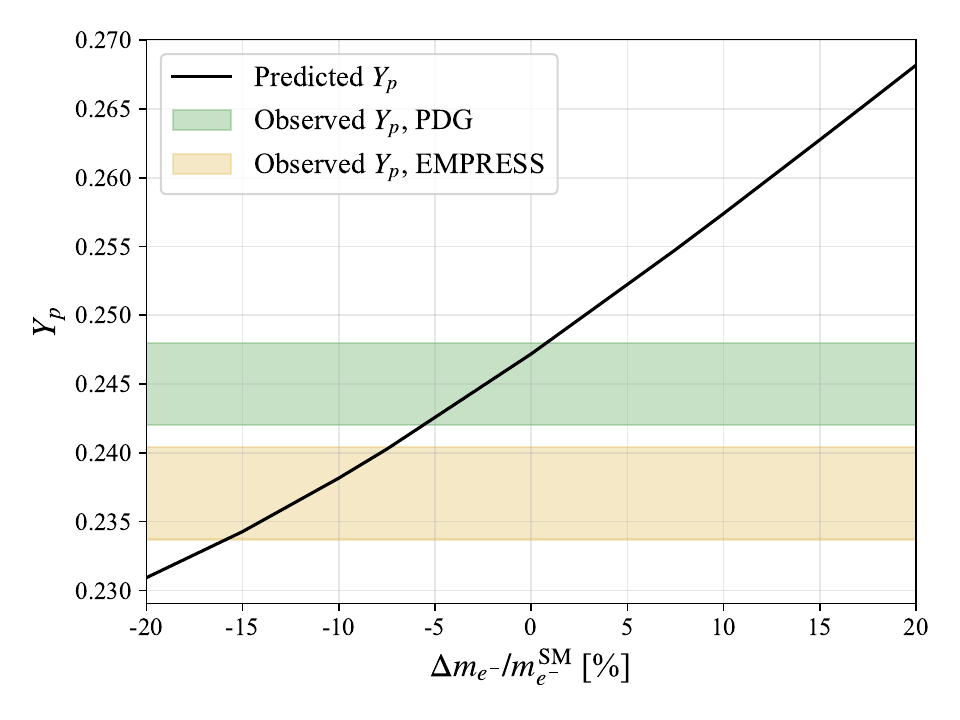}
        \label{fig:1D_plots-a}
    \end{subfigure}\hfill
    \begin{subfigure}{0.48\textwidth}
        \centering
        \includegraphics[width=\linewidth]{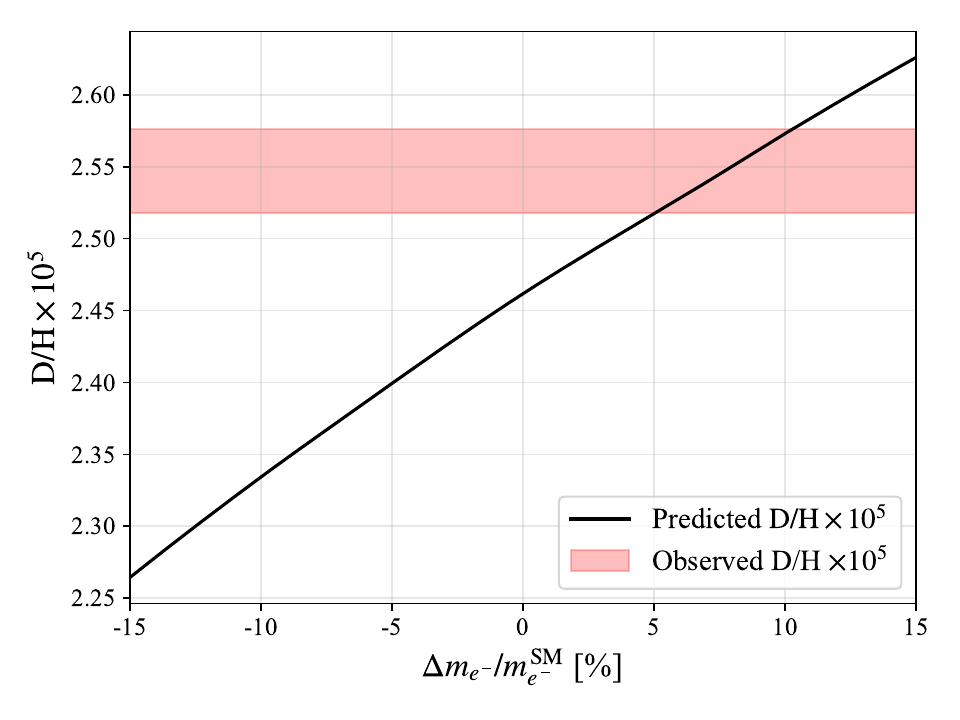}
        \label{fig:1D_plots-b}
    \end{subfigure}

    \medskip

    \begin{subfigure}{0.48\textwidth}
        \centering
        \includegraphics[width=\linewidth]{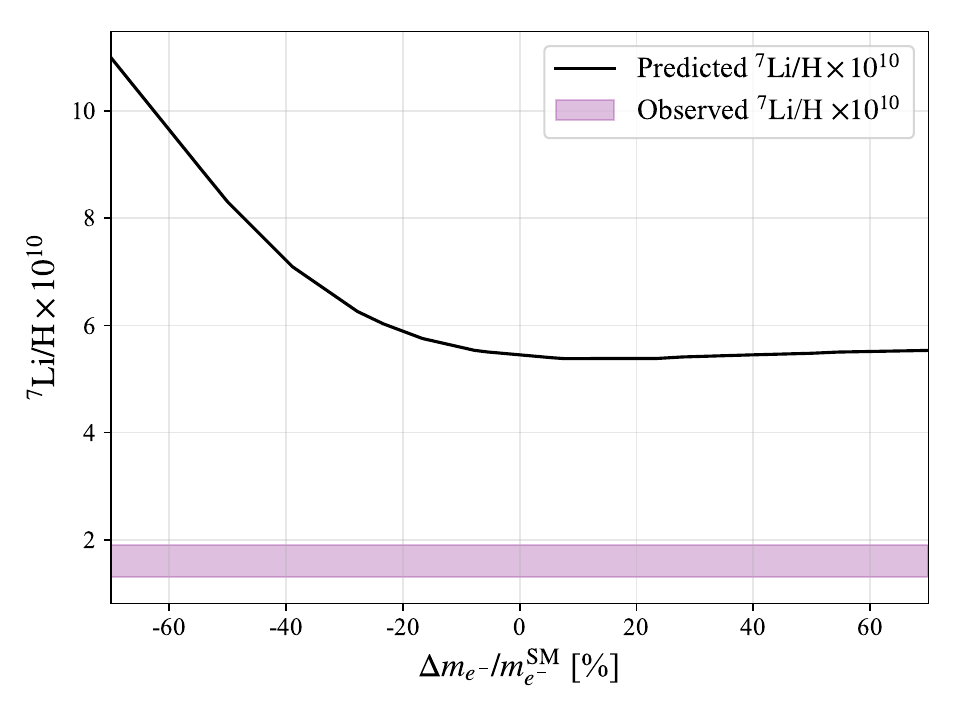}
        \label{fig:1D_plots-c}
    \end{subfigure}\hfill
    \begin{subfigure}{0.48\textwidth}
        \centering
        \includegraphics[width=\linewidth]{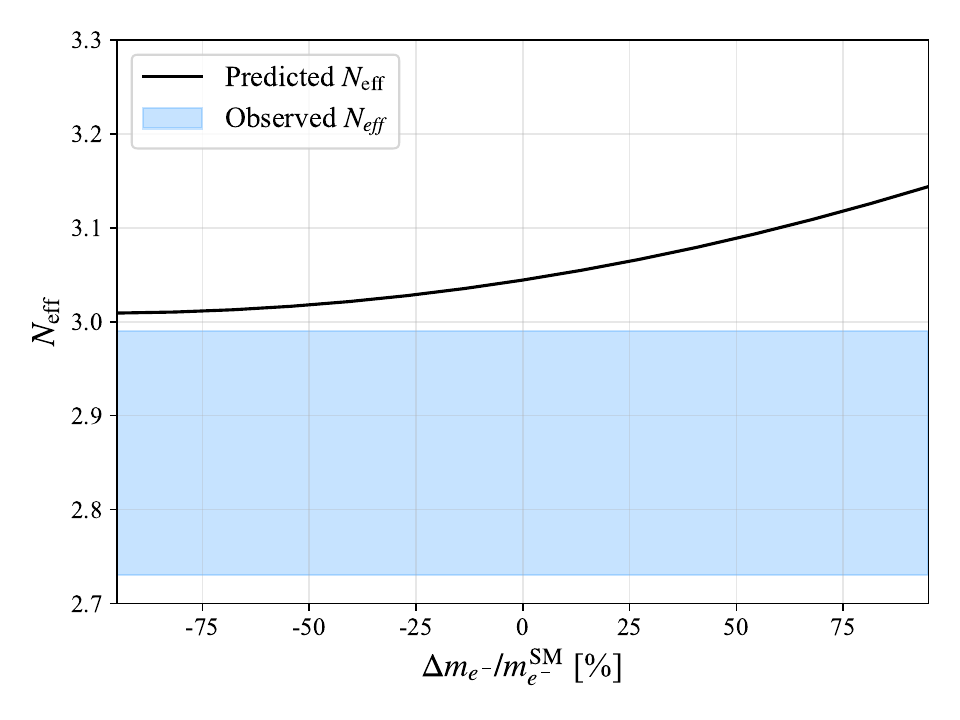}
        \label{fig:1D_plots-d}
    \end{subfigure}

    \caption{Change in the predicted Helium-4, Deuterium, and Lithium-7 abundances as well as $N_{\rm eff}$ with changing electron mass. Here, the electron and positron masses are assumed to be equal. Observational values for Helium-4 come from the PDG and the EMPRESS collaboration \cite{ParticleDataGroup:2024cfk, Matsumoto:2022tlr}. Observation values for Deuterium and Lithium-7 come from the PDG \cite{ParticleDataGroup:2024cfk}. The observed value for $N_{\rm eff}$ comes from ACT \cite{ACT:2025tim}.}
    \label{fig:1D_plots}
\end{figure}

The plots in Figure \ref{fig:1D_plots} show changes in the abundances of Helium-4 ($Y_p$), Deuterium, Lithium-7, and $N_{\rm eff}$ with varied electron mass. These plots were created using \faGithub \href{https://github.com/vallima/PRyMordial}{\,\texttt{PRyMordial's}} full capability for accuracy, but do not distinguish between electron and positron masses.   Observational values for Helium-4 come from the PDG and the EMPRESS collaboration \cite{ParticleDataGroup:2024cfk, Matsumoto:2022tlr}. Observation values for Deuterium and Lithium-7 come from the PDG \cite{ParticleDataGroup:2024cfk}. The observed value for $N_{\rm eff}$ comes from ACT \cite{ACT:2025tim}. Given the persistent tension in the observational determination of Li-7, we treat its inferred abundance cautiously and do not base our primary conclusions on it.

\subsection{Dynamically Solved Chemical Potential}

\begin{figure}[ht]
    \centering
    \includegraphics[width=\linewidth]{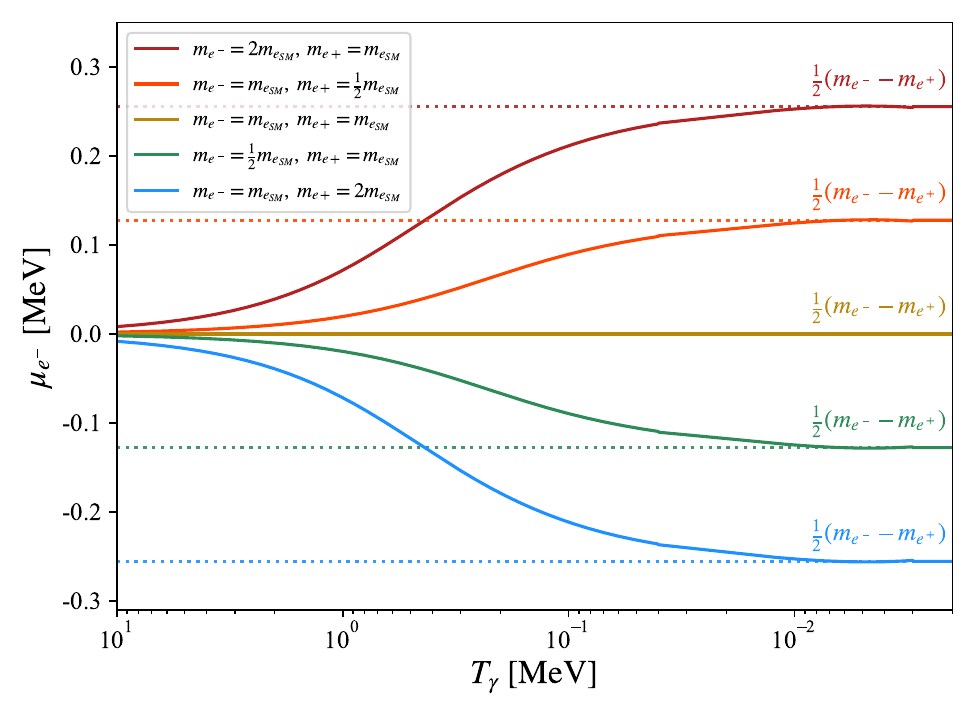}
    \caption{Change in the electron chemical potential over time for several combinations of electron and positron masses. The chemical potential $\mu_{e^-}(T)$ approaches $\frac{1}{2}(m_{e^-} - m_{e^+})$ at low temperature (dotted lines), validating our numerical solution against the analytical limit in Equation \ref{eq:mu_approx}.}
    \label{fig:mu_vs_T}
\end{figure}

Figure \ref{fig:mu_vs_T} shows the change in electron chemical potential as a function of temperature for several combinations of the electron and positron masses. The low temperature value, $\frac{1}{2}(m_{e^-} - m_{e^+})$, is shown on the dotted lines for each case. 

\begin{figure}[ht]
    \centering
    \includegraphics[width=\linewidth]{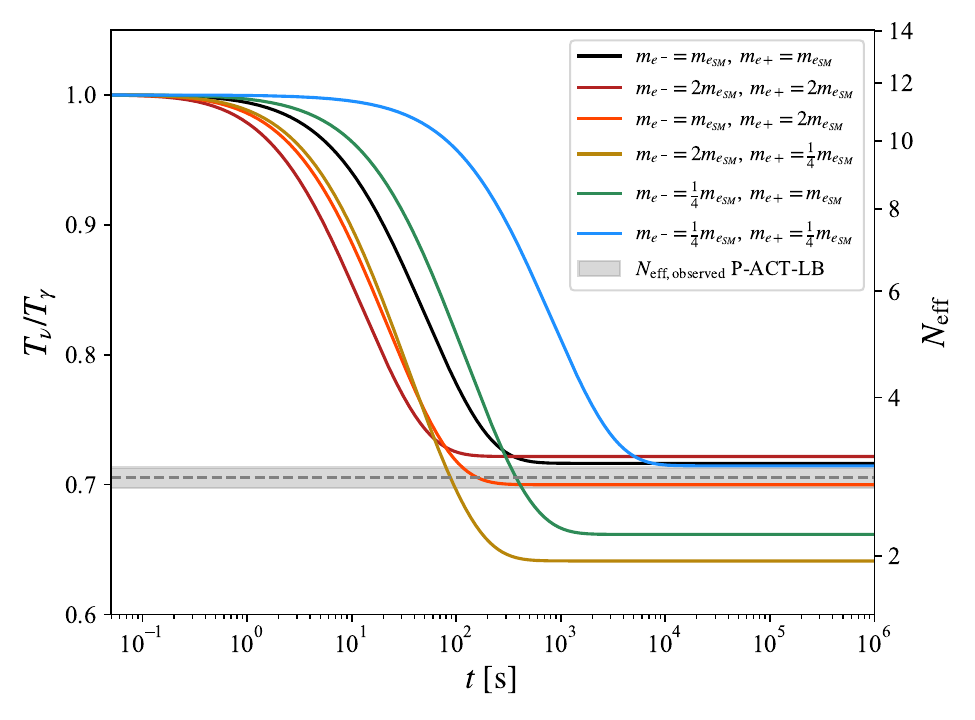}
    \caption{Change in $T_\nu / T_\gamma$ over time for several electron-positron mass differences.}
    \label{fig:T_vs_t}
\end{figure}

For the cases in which their masses are unequal, the electron and positron abundances evolve differently. Imposing a non-zero chemical potential  ensures that charge neutrality is maintained at all times, which in turn shifts the relative populations. The heavier species becomes Boltzmann suppressed earlier, while the lighter one dominates over a range of temperatures. This disparity in the abundances modifies both the scattering and annihilation rates that control neutrino decoupling and the entropy released during annihilation. These features directly determine how entropy is shared between photons and neutrinos, leading to the changes in $T_\nu/T_\gamma$ shown in Figure~\ref{fig:T_vs_t}. The change in neutrino temperature is agnostic to whether the electron mass is changing or the positron mass is changing, i.e. the change in neutrino temperature is the same for $m_{e^-} = 0.5 m_{e^-}^{\rm SM}$, $m_{e^+} = m_{e^-}^{\rm SM}$ and $m_{e^-} = m_{e^-}^{\rm SM}$, $m_{e^+} = 0.5 m_{e^-}^{\rm SM}$.

In the early universe, neutrinos decouple from the thermal bath just before electron-positron annihilation. The subsequent decrease below unity in $T_\nu / T_\gamma$, as shown on the plot, is due to the heating of the photon-bath via transfer of entropy from electrons and positrons to photons during electron-positron annihilation. Lighter combinations of electrons and positrons remain relativistic longer, and therefore annihilate later. Vice-versa, heavier combinations of electrons and positrons annihilate sooner.

The change in the final value of the neutrino temperature is related to the transfer of entropy to the photons from electron positron annihilation. In the case in which $m_{e^+} = m_{e^-} = 2 m_{e,{\rm SM}}$ (red line), electron positron annihilation happens earlier relative to the standard case, during the time at which neutrinos are still partially coupled to the photons. The result of this is that the final value of the ratio of neutrino to photon temperature is higher than the standard value, $T_\nu / T_\gamma = (4/11)^{1/3}$. In the case of both the electron and positron being very light, $m_{e^+} = m_{e^-} = \frac{1}{4} m_{e,{\rm SM}}$ (blue line), the final temperature ratio is lower than the standard case because the neutrinos receive less entropy from electron positron annihilation. There is an interplay here with the decoupling time: lower mass electrons and positrons will annihilate later, but also cause neutrino decoupling to happen later, hence the subtle differences between the high mass (red) and low mass (blue).

In the asymmetric-mass cases (orange, gold, and green curves), the simple intuition based on "annihilation before or after decoupling" is not sufficient. Because charge neutrality fixes a nonzero chemical potential, the relative abundances of electrons and positrons differ, shifting both the scattering rate that controls neutrino decoupling and the entropy release during annihilation in a nontrivial way. As a result, the final value of $T_\nu/T_\gamma$ cannot be anticipated by eye and must be obtained from the full numerical solution.

The grey band on the plot shows the most recent observational value of $N_{eff}$, 2.86 ± 0.13, which has been determined using a combination of data from the Planck legacy CMB data set, Planck CMB lensing and BOSS BAO data, and ACT DR6 \cite{ACT:2025tim}. As shown on the plot, there is a small tension between the SM value of $N_{eff}$ and the latest observed value which can be relieved via the correct choices of the electron and positron masses. 

\subsection{Corrections to the Weak Rates}

The masses of the electron and positron play and essential role in the calculation of the proton-neutron interconversion rates. In this study, the rates were computed from the Born rates with radiative corrections at zero temperature and finite nucleon mass corrections included. These corrections are standard on-shell renormalized radiative corrections to neutron beta decay and only depend on the electron and positron masses through the kinematics of the process. For this reason, we are able to apply these corrections to all of the regions of parameter space in which we are interested. In addition, we incorporate finite-temperature corrections to the weak rates in the plasma, comprising bremsstrahlung contributions and one-loop QED effects. These corrections were derived using the standard electron mass, and are therefore valid only for small deviations of the electron and positron masses away from the SM value. In our analysis we include these corrections only for electron and positron masses within 20\% of the SM mass, and set them to zero otherwise. In the SM like region, including thermal corrections modifies the weak rates by at most 0.1\%, so neglecting them outside of them SM-like region has a negligible impact on our final results \cite{Pitrou:2018cgg}. For details on these corrections, see Refs.~\cite{Burns:2023sgx, Pitrou:2018cgg}.

\subsection{Further Corrections}

The next modification to the standard calculation that is made is to include the effect of variable electron and positron masses on the finite mass corrections to the collision terms in the Boltzmann equation governing the energy densities of the neutrinos. In the unmodified version of the code, finite electron mass effects on the electron-neutrino scattering and annihilation rates are obtained from the updated NUDEC\_BSM\_v2 database \cite{Escudero:2025kej}, which includes recent improvements over the original NUDEC\_BSM calculations \cite{Escudero:2018mvt}. 

To incorporate differing electron and positron masses into the finite-mass corrections that enter the Boltzmann collision terms for neutrino energy densities, we generalize the NUDEC\_BSM\_v2 factors which are provided as functions of the photon temperature. We reinterpret these tables as functions of the dimensionless ratio $m_{e^{-},{\rm SM}} / T$ and then evaluate at the appropriate $m_{e^{\pm}} / T$, thereby assuming that the leading mass dependence of the matrix-element corrections is captured by $m/T$. For scattering off of charged leptons, the finite-mass correction multiplies a process whose rate is proportional to the instantaneous target abundance. We therefore use density-weighted effective factors by averaging the electron and positron corrections with weights given by their instantaneous number densities, which we compute using the dynamically determined electron chemical potential. We then calculate the annihilation terms using a symmetric mean effective factor as electrons and positrons enter symmetrically in this calculation. The Pauli Blocking coefficients were kept unchanged.

In order to correctly account for differing positron and electron mass values in finite-temperature QED corrections to the plasma, we rescale the precomputed interaction pressure and its derivatives, tabulated in \cite{Escudero:2025kej}. For modest changes to the electrons and positron masses such as the ones considered here, we approximate this correction as a function of an effective electron mass, which is chosen to be the average of the electron and positron masses. Using the scaling, $P_{int}(T;m_{e}) = T^4\hat{P}(m_e/T)$, we calculate $P_{int}(T;m_{eff})$ and its derivatives via both an appropriate rescaling of the temperature argument and overall normalization factor of the tabulated rates. Given that QED corrections to the plasma improve the accuracy of the final absolute Helium-4 mass fraction by $\mathcal{O}(10^{-5})$ and the final Deuterium abundance value by $\mathcal{O}(10^{-3})$, this approximation is sufficient for our purposes here. 

Finally, the nuclear reaction rates are normalized using nuclear mass values calculated using atomic mass excess values for neutral atoms from NUBASE2020 \cite{Kondev_2021}. To get the correct nuclear masses used to normalize the nuclear reaction rates in the code, it is necessary to subtract off the contribution from the SM electron mass from mass excess values, as opposed to the modified electron mass.

\subsection{Results}

\begin{figure}[ht]
    \centering
    \begin{subfigure}{0.48\textwidth}
        \centering
        \includegraphics[width=\linewidth]{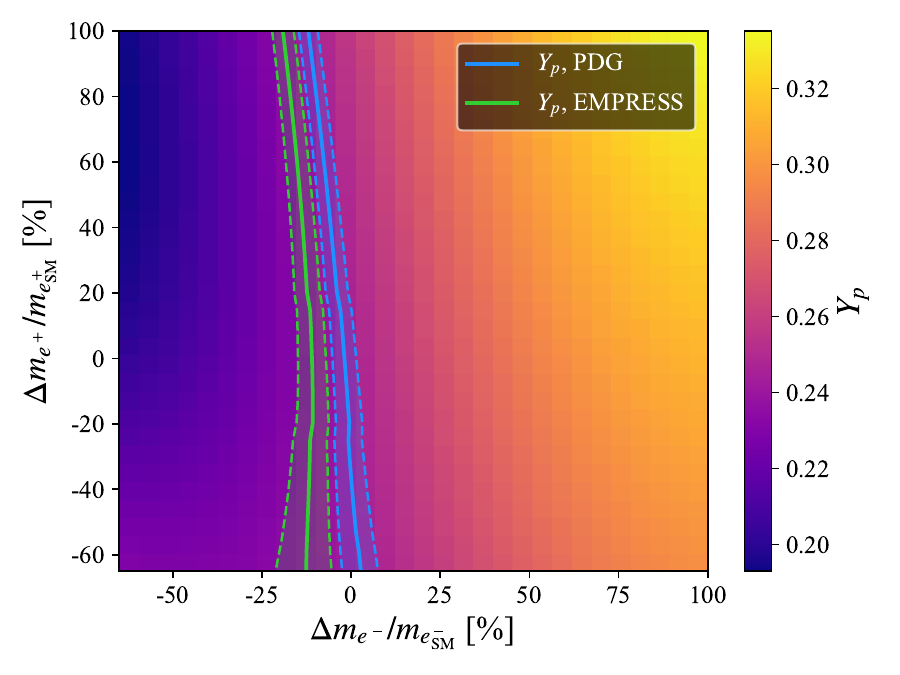}
        \label{fig:heat_maps_3B-a}
    \end{subfigure}\hfill
    \begin{subfigure}{0.48\textwidth}
        \centering
        \includegraphics[width=\linewidth]{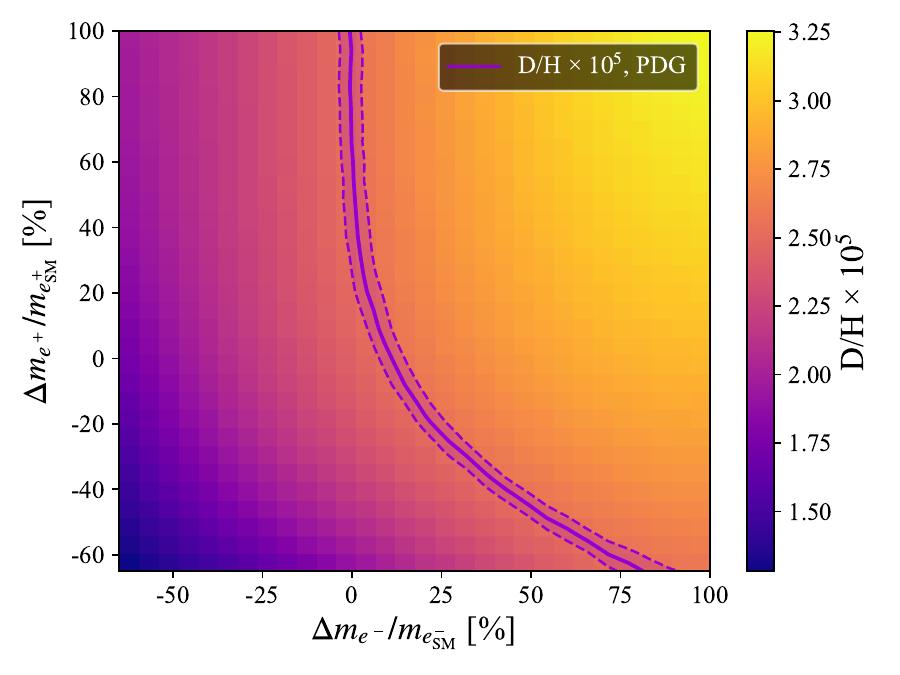}
        \label{fig:heat_maps_3B-b}
    \end{subfigure}

    \medskip

    \begin{subfigure}{0.48\textwidth}
        \centering
        \includegraphics[width=\linewidth]{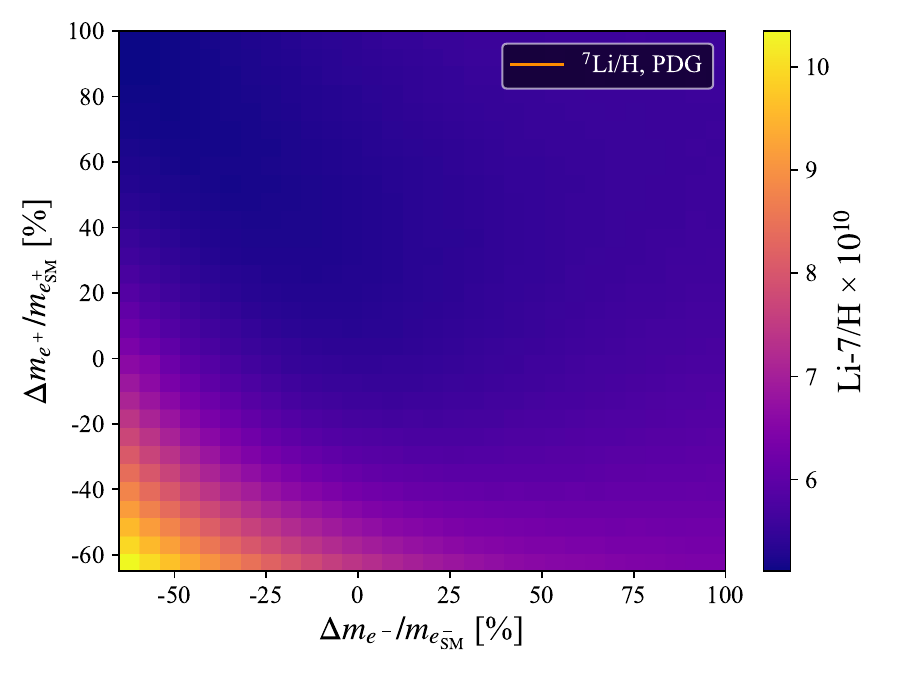}
        \label{fig:heat_maps_3B-d}
    \end{subfigure}
    \begin{subfigure}{0.48\textwidth}
        \centering
        \includegraphics[width=\linewidth]{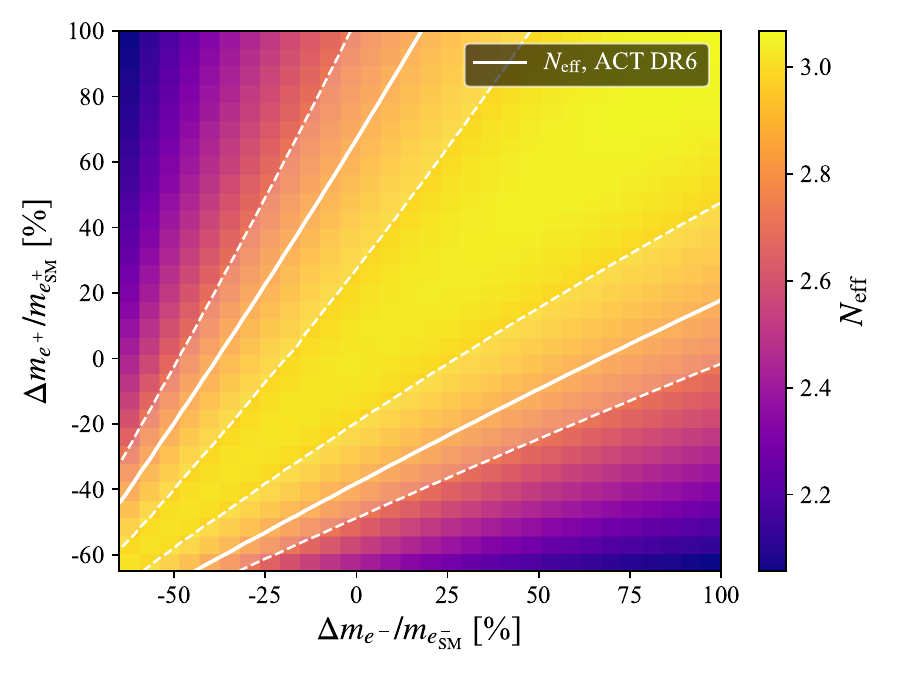}
        \label{fig:heat_maps_3B-d}
    \end{subfigure}

    \caption{Change in the predicted values of Helium-4, Deuterium, Lithium-7, and $N_{\rm eff}$ with changing electron and positron masses. These plots were created under the assumptions outlined above, using dynamically solved non-zero chemical potentials for the electron and positron, the full neutron-proton interconversion rates, and appropriate modifications to the interactions between electrons, positrons, and neutrinos and QED plasma effects. Observational values for Helium-4 come from the PDG and the EMPRESS collaboration \cite{ParticleDataGroup:2024cfk, Matsumoto:2022tlr}. Observation values for Deuterium and Lithium-7 come from the PDG \cite{ParticleDataGroup:2024cfk}. The observed value for $N_{\rm eff}$ comes from ACT \cite{ACT:2025tim}.}
    \label{fig:heat_maps_3B}
\end{figure}

The plots in Figure \ref{fig:heat_maps_3B} are heat maps showing the resulting values of the Helium-4 abundance, Deuterium abundance, Lithium-7 abundance, and $N_{eff}$ for varying values of the electron and positron masses. For these ranges of electron and positron masses, the predicted Lithium-7 value does not reach the observed value, and thus the contours for the observed value are not visible on the plot. We do not consider this to be problematic for our results as it has become clear in recent years that more observational data for the Lithium-7 abundance in stars is needed in order for the values to be taken seriously in comparison to predicted values \cite{Fields:2022mpw}.

\begin{figure}[ht]
    \centering
    \begin{subfigure}{0.48\textwidth}
        \centering
        \includegraphics[width=\linewidth]{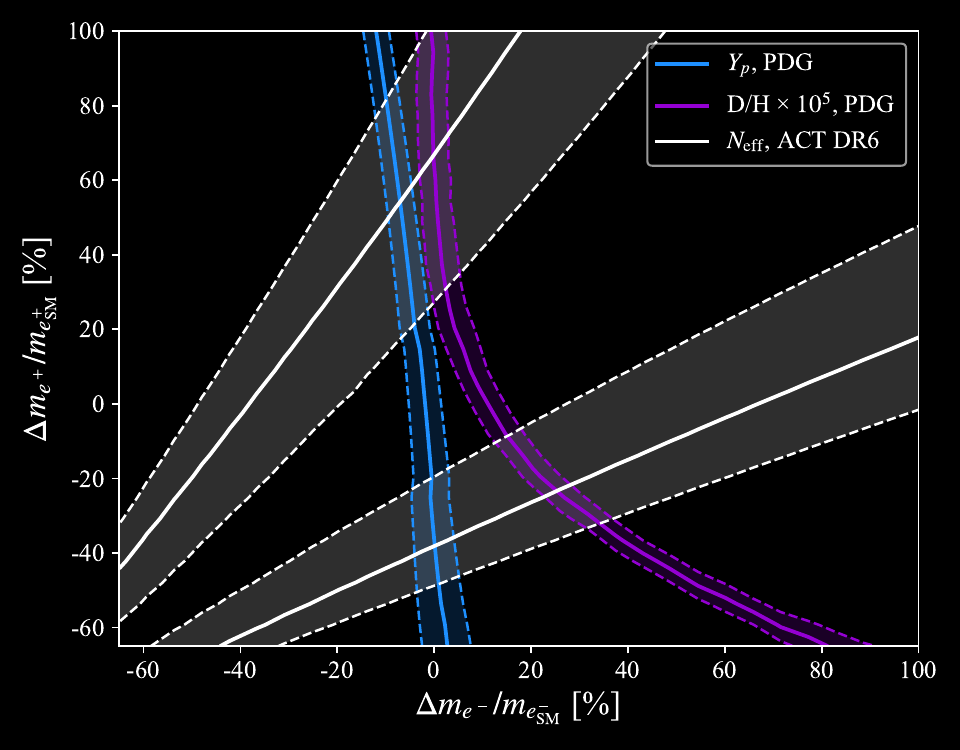}
        \label{fig:overlap-e}
    \end{subfigure}\hfill
    \begin{subfigure}{0.48\textwidth}
        \centering
        \includegraphics[width=\linewidth]{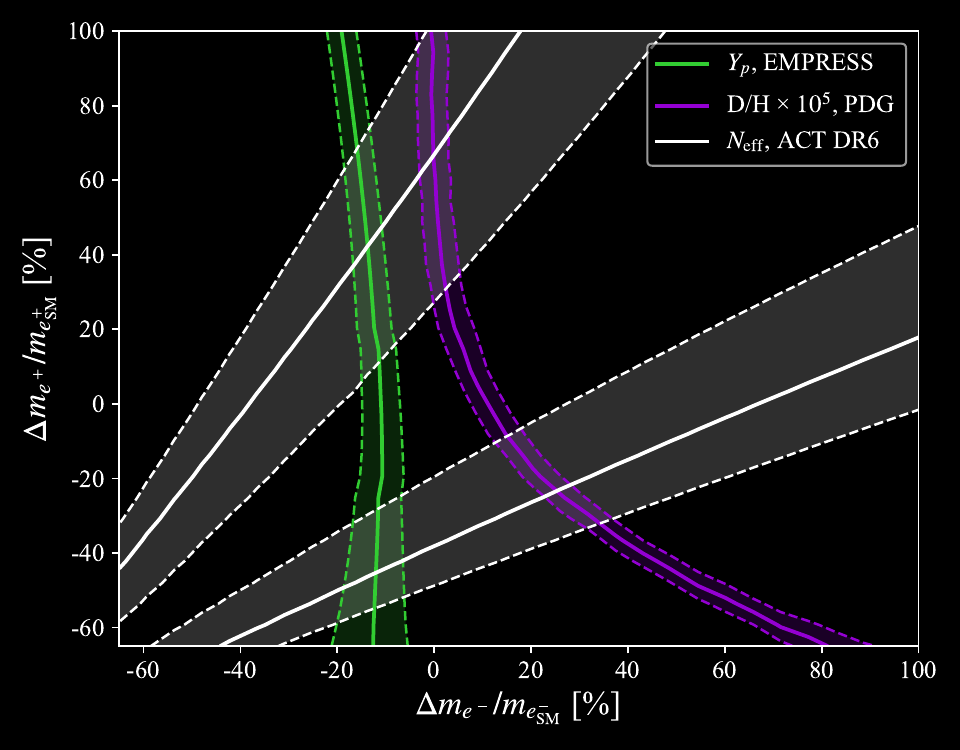}
        \label{fig:overlap-f}
    \end{subfigure}

    \caption{Overlap regions in ($m_{e^-}, m_{e^+}$) parameter space of the bounds from observation from Helium-4, Deuterium, and $N_{\rm eff}$. The plot on the left uses the EMPRESS value for Helium-4, and the plot on the right uses the PDG value. These plots were created under the assumptions outlined above, using dynamically solved non-zero chemical potentials for the electron and positron, the full neutron-proton interconversion rates, and appropriate modifications to the interactions between electrons, positrons, and neutrinos and QED plasma effects. Observational values for Helium-4 come from the PDG and the EMPRESS collaboration \cite{ParticleDataGroup:2024cfk, Matsumoto:2022tlr}. Observation values for Deuterium come from the PDG \cite{ParticleDataGroup:2024cfk}. The observed value for $N_{\rm eff}$ comes from ACT \cite{ACT:2025tim}.}
    \label{fig:overlap}
\end{figure}

Figure \ref{fig:overlap} shows the regions of overlap of the bounds from observation from Helium-4, Deuterium, and $N_{eff}$ in the same range of parameter space as is used for the plots in Figure \ref{fig:heat_maps_3B}. For clarity, the observational contours from the two different sets of Helium-4 data are shown on different plots. The plot on the left uses the EMPRESS value for Helium-4, and the plot on the right uses the PDG value.

As is shown in Figure \ref{fig:overlap}, there is no combination of electron and positron masses for which all three observations align at the level of 1$\sigma$. That said, a particularly interesting region emerges when $m_{e^-}$
is reduced by 1-4\% relative to its SM value while $m_{e^+}$ is increased by 25-60\% : in this slice of parameter space, the three observations approach a 1$\sigma$ overlap when the PDG Helium-4 value is adopted. Using the EMPRESS Helium-4 determination, the contours overlap only at the 3$\sigma$ level.

\section{Theoretical Mechanisms for Temperature-Dependent CPT Violation}
\label{sec:theory_models}

Having established the BBN constraints on electron-positron mass
asymmetries, we now discuss several theoretical frameworks that could
naturally generate temperature-dependent CPT violation. The key
requirement is a mechanism that produces $b_0(T) \propto T^2$ scaling,
enabling significant CPT violation in the early universe while
automatically satisfying stringent present-day experimental bounds.

Recent work by Baryakhtar \textit{et al.}\ has emphasized that
cosmological scalar fields coupled to the Standard Model can induce
early-time or redshift-enhanced temporal variations in quantities such
as the electron mass and fine-structure constant, reaffirming that
cosmological epochs are powerful probes of such
couplings~\cite{Baryakhtar:2025uxs,Baryakhtar:2024rky}.

We present two complementary toy models below, each demonstrating
different field-theoretic mechanisms that can produce the desired
temperature dependence. The
first model (Section~\ref{sec:scalar_vector}) 
derives the $T^2$ scaling from higher-dimensional operators and
thermal scalar condensates through a temperature-driven phase
transition, providing a natural UV structure. The second model
(Section~\ref{sec:pt_symmetric}) explores a non-Hermitian
PT-symmetric approach, showing that alternative
field-theoretic frameworks can also generate the
required behavior. In standard effective field theories in which spontaneous symmetry breaking occurs, symmetry generally is broken at low temperature and restored at high temperature. Here, our models show the opposite behavior, known as inverse symmetry breaking in the literature. In his 1974 paper Weinberg was the first to note the possibility of symmetry non-restoration \cite{Weinberg:1974hy}. This idea was further explored by Mohapatra and Senjanović in 1979 \cite{Mohapatra:1979qt,Mohapatra:1979vr} and Dvali and Senjanović in 1995 \cite{Dvali:1995cc}.

Together, these examples serve as existence proofs
that the mass asymmetries constrained by BBN can be realized in
theoretically reasonable frameworks, though a complete UV-complete
theory remains an open question for future work.


\subsection{Connecting the \texorpdfstring{$a_\mu$}{a-mu} Framework
            to the Mass-Splitting Parametrization}
\label{sec:bmu_connection}
The BBN analysis of Section~\ref{sec:BBN_constraints} is cast in
terms of a direct mass difference $\delta m \equiv m_{e^+} - m_{e^-}$,
while the toy models below generate CPT violation and subsequent mass shifts through a vector
background field $B_\mu$ coupled to the fermion current.  It is
important to establish the quantitative relationship between the two
parametrizations so that the BBN bounds can be interpreted in the
language of the Standard Model Extension
(SME)~\cite{Colladay:1998fq,Kostelecky:2008ts}.
\paragraph*{Dispersion relations.}
We couple $B_\mu$ to the fermion via the CPT-odd vector operator
$-g B_\mu\bar\psi\gamma^\mu\psi$ (see Section~\ref{sec:scalar_vector} below), so that the temporal VEV
$a_0 \equiv g\langle B_0\rangle$ plays the role of the SME
coefficient $a_0^e$.  For a purely temporal background
$a^\mu = (a_0,\mathbf{0})$, the modified Dirac equation
$(\slashed{p} - m - a_0\gamma^0)u = 0$ yields the exact dispersion
relation~\cite{Colladay:1998fq}
\begin{equation}
    E^{e^\mp}(\mathbf{p})
    = \sqrt{\mathbf{p}^2 + m^2} \pm a_0 ,
    \label{eq:vector_dispersion}
\end{equation}
where the upper (lower) sign applies to the electron (positron).
The opposite sign for the positron follows from charge conjugation,
which maps $a_\mu \to -a_\mu$ for the antiparticle. The
axial-vector operator, $b_\mu\bar\psi\gamma^\mu\gamma^5\psi$, shifts the momentum argument of the dispersion relation, $E= \sqrt{m^2+(p \pm b_0)^2}$ producing a spin-dependent effect that vanishes at rest. In addition, being C-even, it shifts particles and antiparticles identically and therefore generates no particle--antiparticle mass splitting. Conversely, the vector operator shifts the energies of
electrons and positrons in \emph{opposite} directions at all momenta independently of spin.

\paragraph*{Effective mass splitting.}
At $\mathbf{p} = 0$, Eq.~\eqref{eq:vector_dispersion} gives rest
energies $m + a_0$ for the electron and $m - a_0$ for the positron,
yielding the exact effective mass splitting
\begin{equation}
    \delta m_{\rm eff}
    \equiv E^{e^+}(0) - E^{e^-}(0)
    = -2a_0 ,
    \qquad
    |\delta m_{\rm eff}| = 2a_0 .
    \label{eq:dm_eff}
\end{equation}
This identification is exact and linear in $a_0$, with no
spin-averaging subtlety: the vector coupling shifts all spin states of
the electron (positron) up (down) by the same amount $a_0$.  The
vector operator therefore acts in the Fermi-Dirac distributions
exactly as a shift of the chemical potential by $\mp a_0$ for
electrons and positrons respectively---equivalent, to leading order,
to a rest-mass splitting $\delta m = 2a_0$.
\paragraph*{Mapping to BBN constraints.}
Since $a_0(T) = g\langle B_0(T)\rangle \propto T^2$ in both
models below, we have
\begin{equation}
    \delta m_{\rm eff}(T) = 2\,g\langle B_0(T)\rangle
    = 2\,a_0(T) \propto T^2 ,
    \label{eq:b0_dm_map}
\end{equation}
connecting the toy-model parameter $\alpha$ (defined by
$b_0(T) = \alpha T^2$ in the phenomenological analysis) to the
underlying coupling via $\alpha = g \xi c_\phi /(\lambda_\phi m_B^2)$ in Models~I and~II.  The BBN bounds derived in
Section~\ref{sec:BBN_constraints} therefore constrain
$a_0(T_{\rm BBN}) \lesssim \mathcal{O}(\mathrm{keV})$, representing
early-universe constraints on the SME vector coefficient $a_e^\mu$
complementary to the laboratory bounds in the \textit{Data Tables for
Lorentz and CPT Violation}~\cite{Kostelecky:2008ts}.

The identification $\delta m \approx 2a_0^e$ is justified by the fact that $a_\mu$ is species specific, along with the
kinematics of the $n$--$p$ interconversion reactions themselves.
In our model $a_\mu$ couples only to
electrons and positrons, with $a_\mu^\nu = 0$ for neutrinos. The
physically relevant quantity is therefore $\Delta a_0 = a_0^e - a_0^\nu
= a_0^e \neq 0$. This is observable because
electrons and neutrinos share interaction terms---specifically, the
charged-current $W$-exchange vertices, $W_\mu\,\bar\psi_{\nu_e}\gamma^\mu\psi_{e^-}$, 
that govern the $n$--$p$ interconversion reactions central to BBN. The field redefinition that removes $a_0^e$ from the electron kinetic term cannot simultaneously remove it from the charged-current vertex without introducing it into the neutrino sector, so the relative shift $\Delta a_0$ is physical. Physically, $\Delta a_0 = a_0^e$ shifts the effective
energy thresholds of these weak processes, modifying the neutron-to-proton
ratio at freeze-out and hence the primordial abundances, as discussed in previous sections.

Additionally, 
at $T_{\rm BBN} \sim 1\,\mathrm{MeV}$, with $m_e \simeq 0.511\,\mathrm{MeV}$,
the thermal electron-positron plasma is mildly relativistic, so one
cannot appeal to a blanket non-relativistic approximation.
However, the weak-rate phase-space integrals for processes such as
$n + e^+ \to p + \bar\nu_e$ and $n \to p + e^- + \bar\nu_e$
are weighted most heavily at the kinematic threshold of each reaction,
where the lepton three-momentum $|\mathbf{p}| \to 0$.
At threshold, Eq.~\eqref{eq:vector_dispersion} reduces exactly to
$E^{e^\mp}(0) = m \pm a_0$, so the rest-energy splitting
$\delta m_{\rm eff} = 2a_0$ is not an approximation in the
threshold-dominated region but an exact result.
The difference between the $a_0$ and mass-difference dispersion
relations is of order $a_0\,(1 - m/E)$, which vanishes at threshold
and is suppressed relative to the leading effect across the
entire phase-space integral.
The identification $\delta m \approx 2a_0^e$ therefore captures
the dominant BBN physics without requiring the plasma to be
non-relativistic as a whole.
\subsection{Natural Toy Model: Scalar-Vector Coupling}
\label{sec:scalar_vector}

Here we consider a mechanism where
$b_0(T) \propto T^2$ arises dynamically from a temperature-driven
phase transition and scalar-vector interactions, without requiring
ad-hoc terms in the potential.

To generate a CPT-violating background with natural $T^2$ scaling, we
introduce a scalar field $\phi$ and a vector field $B_\mu$, with an interaction that couples $\phi^2$ linearly to
$B_\mu$ via the thermal rest-frame vector $u^\mu = (1, 0, 0, 0)$.
The Lagrangian is:
\begin{equation}
\mathcal{L} = -\frac{1}{4} F_{\mu\nu}F^{\mu\nu}
+ \frac{1}{2} (\partial_\mu \phi)^2 - V_\phi(\phi,T)
+ \bar{\psi}\!\left(i \slashed{\partial} - m
- g B_\mu \gamma^\mu\right)\!\psi
- \frac{\lambda_B}{4} (B_\mu B^\mu)^2
- \xi \phi^2\, u_\mu B^\mu
\label{eq:L_modelII}
\end{equation}
where:
\begin{equation}
V_\phi(\phi,T) = \frac{\mu_\phi^2}{2} \phi^2
+ \frac{\lambda_\phi}{4} \phi^4
- \frac{1}{2} c_\phi T^2 \phi^2
\end{equation}

The dimension-4 operator $\xi \phi^2\, u_\mu B^\mu$ is
the key ingredient, where $\xi$ is a dimension-1 coupling. The coupling $\xi$ is radiatively stable: at $\xi = 0$ the scalar $\phi$ decouples entirely from the
$B_\mu$--$\psi$ sector, so no loop diagram can generate this
operator; all corrections are therefore proportional to $\xi$ itself
($\delta\xi \propto \xi$), making $\xi = 0$ a technically natural
point. This term explicitly breaks Lorentz
invariance in the thermal rest frame---a feature that is physically
acceptable since finite temperature already singles out a preferred
frame. Similar operators appear naturally in effective field theories
of Lorentz violation and in models with spontaneous breaking of
spacetime symmetries~\cite{Colladay:1998fq}. The thermal rest-frame
vector $u^\mu$ can be understood as arising from the presence of the
thermal bath itself, making this construction phenomenologically
well-motivated in cosmological contexts.
The vector coupling (without $\gamma^5$)
ensures that the temporal VEV $a_0 = g\langle B_0\rangle$ produces
a direct mass splitting $\delta m_{\rm eff} = 2a_0$ between electrons
and positrons.

The scalar potential is constructed so that at zero temperature,
$\phi = 0$ is the stable minimum (since $\mu_\phi^2 > 0$), but at
high temperatures the negative thermal mass term
$-\frac{1}{2} c_\phi T^2 \phi^2$ can drive inverse symmetry
breaking~\cite{Weinberg:1974hy,Mohapatra:1979qt,Dvali:1995cc}. Minimizing the potential:
\begin{equation}
\frac{\partial V_\phi}{\partial \phi}
= \mu_\phi^2 \phi + \lambda_\phi \phi^3 - c_\phi T^2 \phi = 0
\end{equation}
For $\phi \neq 0$:
\begin{equation}
\langle \phi(T) \rangle
= \sqrt{\frac{c_\phi T^2 - \mu_\phi^2}{\lambda_\phi}}
\end{equation}

This is only real and nonzero when $T > T_c \equiv \mu_\phi/\sqrt{c_\phi}$.
For $T < T_c$, we have $\langle\phi\rangle = 0$.  Thus at high
temperatures $T \gg T_c$, the scalar acquires a thermal vacuum
expectation value: $\langle \phi(T) \rangle \sim T$, while at low
temperatures (including $T = 0$), $\langle\phi\rangle = 0$, ensuring
$b_0(T) \to 0$ as required.

The interaction $\xi \phi^2 u_\mu B^\mu$ induces a
temperature-dependent VEV for the time component of the vector field.
In the regime where the quartic self-coupling $\lambda_B$ is negligible
compared to an effective mass term (which we denote $m_B^2$), the
effective potential for $B_0$ becomes:
\begin{equation}
V_{\rm eff}(B_0) = \frac{1}{2} m_B^2 B_0^2
- \xi \langle \phi(T) \rangle^2 B_0
\end{equation}
Minimizing:
\begin{equation}
\frac{\partial V_{\rm eff}}{\partial B_0}
= m_B^2 B_0 - \xi \langle \phi(T) \rangle^2 = 0
\quad \Rightarrow \quad
\langle B_0(T) \rangle
= \frac{\xi}{m_B^2} \langle \phi(T) \rangle^2
\end{equation}

For $T \gg T_c$ where $\langle \phi(T) \rangle \sim T$, this yields
$\langle B_0(T) \rangle \sim T^2$, and hence
$a_0(T) = g \langle B_0(T) \rangle \propto T^2$.
Crucially, for $T < T_c$ (including $T = 0$), $\langle\phi\rangle = 0$
and therefore $a_0(T) = 0$, satisfying experimental bounds. The
critical temperature $T_c$ can be chosen to be well below BBN
temperatures---for example, $T_c \sim$ keV gives
$T_{\rm BBN}/T_c \sim 10^3$, ensuring we are in the $T \gg T_c$
regime at BBN---while still being many orders of magnitude above
present-day cosmic temperatures ($T_0 \sim 10^{-4}$ eV).

In this model, the temperature dependence emerges from the
well-understood thermal behavior of scalar fields and a
temperature-driven phase transition. Above the critical temperature $T_c$, the scalar field
develops a nonzero VEV that grows with temperature, inducing the
CPT-violating background. Below $T_c$, the scalar returns to its
symmetric phase with $\langle\phi\rangle = 0$, automatically restoring
CPT symmetry. Observable effects at BBN temperatures can be produced through
appropriate choices of the couplings $\xi$,
$\lambda_\phi$ and $m_B$, and the critical temperature scale $T_c$.
The model does not require
fine-tuning to maintain the $T^2$ scaling against radiative
corrections---the scaling is protected by the structure of the thermal
scalar VEV above the phase transition.

\subsection{PT-Symmetric Toy Model}
\label{sec:pt_symmetric}

As an alternative to the Hermitian vector potential, one can consider
the field $B_0$ treated as a zero-dimensional
(i.e.\ spatially homogeneous) quantum field governed by a
PT-symmetric Hamiltonian. PT-symmetric Hamiltonians are generally
non-Hermitian but invariant under the combined action of parity ($P$)
and time-reversal ($T$) transformations, i.e.\ $[H,PT]=0$.
Remarkably, such Hamiltonians can possess entirely real spectra in the
``unbroken PT phase'' despite being non-Hermitian, allowing for
physically meaningful behavior.

PT-symmetric quantum field theory has been developed systematically in
recent years~\cite{Bender:1998ke,Bender:2004sa,Bender:2007nj,%
Bender:2018pbv,Bender:2005hf,Jones:2009qs}, with rigorous results
establishing the reality of the spectrum, the unitarity of time
evolution (with respect to a modified inner product), and consistent
perturbation theory for a broad class of non-Hermitian actions.  The
framework has been extended to scalar field theories in $d$ spacetime
dimensions, to fermionic systems, and more recently to cosmological
backgrounds.  The key insight is that Hermiticity is sufficient but
not necessary for a consistent quantum theory---what matters is that
the spectrum of physical observables is real.  In PT-symmetric
theories this reality is guaranteed by the unbroken PT symmetry
itself.

In the present context we exploit the $(0+1)$-dimensional reduction of
this framework: $B_0$ is promoted to a \emph{quantum field} that is
spatially uniform but dynamical in the imaginary-time (Euclidean)
direction.  The thermal partition function
\begin{equation}
Z(T) = \mathrm{Tr}\,e^{-H(B_0)/T}
     = \oint\!\mathcal{D}B_0\;
       \exp\!\left[-\int_0^{1/T}\!d\tau\;
       \mathcal{L}_E(B_0,\dot{B}_0)\right],
\label{eq:Z_PT}
\end{equation}
where $\mathcal{L}_E = \frac{1}{2}\dot{B}_0^2 + V(B_0)$ and the path
integral is over field configurations periodic in $\tau$ with period
$1/T$, is precisely the standard finite-temperature path integral of a
$(0+1)$-dimensional quantum field
theory~\cite{Kapusta:2006pm,Bellac:2011kqa}.  The trace in
Eq.~\eqref{eq:Z_PT} is therefore a \emph{thermodynamic} statement
about the full Hilbert space---not a single-particle matrix element.

In this framework, the PT-symmetric Hamiltonian
operator takes the form
\begin{equation}
H(B_0) = \frac{p^2}{2} - i\lambda B_0^3
+ \frac{1}{2}cT^2 B_0^2,
\label{eq:H_PT}
\end{equation}
where $p$ is the momentum conjugate to $B_0$ , and
$\lambda, c > 0$ are real coupling constants.  The cubic interaction
carries an imaginary coefficient, which is \emph{required} by PT
symmetry: under $P: B_0\to -B_0$ and $T: i \to -i$ the term
$-i\lambda B_0^3$ maps to itself, leaving the Hamiltonian invariant.
Thus the imaginary cubic term is the minimal PT-symmetric deformation
of the harmonic oscillator, and its appearance here is structurally
natural~\cite{Bender:1998ke}.
At zero temperature the quadratic term in Eq.~\eqref{eq:H_PT}
vanishes, leaving only the cubic interaction; the
ground state of the purely cubic $\mathcal{PT}$-invariant theory is
known to be well-defined with a real,
spectrum~\cite{Bender:1998ke,Bender:2004sa}.
The ground-state expectation value satisfies
$\langle B_0(0)\rangle = 0$, ensuring CPT symmetry is restored as $T\to 0$. This vanishing is guaranteed by PT symmetry itself: since $B_0$ is odd under combined PT transformation, an unbroken PT-symmetric ground state $|0\rangle$
satisfies $\langle 0|B_0|0\rangle = 0$ exactly.

The thermal effective potential obtained from the
$(0+1)$-dimensional path integral
\begin{equation}
V_{\rm eff}(B_0,T) = -T\ln\mathrm{Tr}\,e^{-H(B_0)/T}
\end{equation}
can then be minimized to give a temperature-dependent
VEV~\cite{Kapusta:2006pm}.
In the high-temperature limit where the quadratic term dominates, a
saddle-point evaluation of the path integral yields
$\langle B_0(T)\rangle\sim T^2$ from the competition between the
thermal quadratic term and the cubic interaction, reproducing the
$T^2$ scaling obtained in the Hermitian model above, but without
requiring an ad-hoc real cubic term in the potential. \footnote{The non-trivial saddle at $B_0^* = -icT^2/(3\lambda)$ is
purely imaginary. In PT-symmetric QFT, physical observables are
computed using the $\mathcal{C}$-inner product~\cite{Bender:2004sa},
under which $\langle B_0\rangle_{\mathcal{C}}$ is real, yielding
$\langle B_0\rangle_{\mathcal{C}} = cT^2/(3\lambda) \propto T^2$.}

We stress that the nonzero $\langle B_0(T)\rangle$ at high temperature
is a collective thermal effect, arising from the competition between
the entropy of fluctuations---encoded in the full trace over the
Hilbert space in Eq.~\eqref{eq:Z_PT}---and the energy cost of
displacing $B_0$ from the origin.  We emphasise, however, that in
$(0+1)$ dimensions the Gibbs state $\rho \propto e^{-H/T}$ is unique
for any finite system, so what occurs here is \emph{not} spontaneous
symmetry breaking in the QFT sense (which requires an infinite-volume
thermodynamic limit).  Rather, the temperature-dependent shift of
$\langle B_0(T)\rangle$ away from zero is \emph{formally analogous to
SSB at the level of the saddle-point structure of the thermal effective
action}: the dominant saddle of the path integral shifts away from
$B_0=0$ at sufficiently high $T$, generating the desired VEV without
requiring true vacuum degeneracy.  The PT-Symmetric model is therefore intended as
a proof of existence of a consistent non-Hermitian thermal saddle, not
as a complete or realistic UV completion.  The extension to a full
$(3+1)$-dimensional non-Hermitian scalar or vector field
theory---where genuine SSB can occur---is actively being developed in
the PT-symmetric QFT
literature~\cite{Bender:2018pbv,Bender:2005hf,Bender:2007nj}.

The PT-symmetric approach has both advantages and limitations. On the
positive side, it provides a mathematically consistent,
rigorously studied alternative to Hermitian field
theory and naturally accommodates an otherwise challenging cubic potential structure. The imaginary coefficient of the cubic
term is required by PT symmetry rather than being
ad-hoc, and the reality of the resulting spectrum in
the unbroken phase has been established to all orders in perturbation
theory~\cite{Bender:2004sa,Jones:2009qs}. On the other hand,
while the $(0+1)$-dimensional QFT formulation is fully
rigorous, the embedding of this mechanism into a complete
$(3+1)$-dimensional non-Hermitian quantum field theory coupled to the
Standard Model fields remains an active area of research, and the
physical interpretation of such couplings requires careful
consideration. Nevertheless, this $(0+1)$-dimensional
QFT example demonstrates that non-standard
field-theoretic frameworks can provide viable
mechanisms for temperature-dependent CPT violation.

This illustrates how PT-symmetric field systems can
provide an alternative mechanism to achieve the desired
temperature-dependent behavior in a more ``allowed'' non-Hermitian
setting, realizing the $T^2$ scaling at the level of the thermal
saddle-point structure.

In summary, both models demonstrate that $b_0(T)\propto T^2$
scaling can arise from well-motivated field-theoretic structures.
The Scalar-Vector coupling model with a temperature-driven phase
transition appears most natural from an effective field theory
perspective, though it requires careful tuning of the critical
temperature scale. The PT-symmetric $(0+1)$-dimensional
QFT model offers an intriguing alternative grounded
in a well-developed theoretical framework, realizing the $T^2$ scaling
at the level of the thermal saddle-point structure rather than through
genuine SSB. Future work should focus on embedding these toy models
into more complete UV theories and exploring their phenomenological
signatures beyond BBN.

\subsection{Remark on Baryogenesis Applications}
\label{sec:baryogenesis}

While the primary focus of this work is on constraining
temperature-dependent CPT violation through BBN observations, we
briefly note that the same framework with much smaller coupling
constants could in principle address baryogenesis while evading some of the standard Sakharov conditions \cite{Bertolami:1996cq}. For
$\alpha \sim 10^{-10}~\text{GeV}^{-1}$, the CPT-violating background
$b_0(T) = \alpha T^2$ would induce negligible mass differences at BBN
temperatures ($\lesssim 10^{-7}$ eV), completely evading all
observational constraints discussed in
Section~\ref{sec:BBN_constraints}. However, at the electroweak phase
transition ($T_{\rm EW} \sim 100$ GeV), this background acts as an
effective chemical potential for lepton number, generating a net lepton
asymmetry that can be partially converted to baryon asymmetry via
sphaleron processes~\cite{Khlebnikov:1988sr, Harvey:1990qw}. A simple
thermal equilibrium estimate yields $\eta_B \sim \alpha T_{\rm EW}$,
suggesting that the observed baryon-to-photon ratio
$\eta_B^{\rm obs} \sim 10^{-10}$ could be reproduced with
$\alpha \sim 2.6 \times 10^{-10}~\text{GeV}^{-1}$~\cite{ParticleDataGroup:2024cfk}.
At today's temperature $T_0 \sim 10^{-13}$ GeV, this gives
$b_0(T_0) \sim 10^{-36}$ GeV, far below laboratory bounds
($\lesssim 10^{-26}$ GeV)~\cite{Kostelecky:2008ts}. This baryogenesis
scenario is fundamentally distinct from the large-$\alpha$ regime
constrained by BBN: the keV-scale mass differences allowed by our BBN
analysis would overproduce the baryon asymmetry by factors of $10^6$
or more, requiring implausibly strong washout mechanisms. Thus, while
the $b_0(T) \propto T^2$ framework can accommodate both applications,
they correspond to vastly different parameter regimes and cannot be
simultaneously realized. A detailed exploration of the baryogenesis
mechanism, including sphaleron conversion dynamics and washout effects, is left for future work.

\section{Assessment of Astrophysical Environments}

We now examine the potential impact of a CPT-violating mass difference between electrons and positrons, $\delta m = m_{e^+} - m_e \sim \text{few keV}$, in astrophysical systems with temperatures in the MeV range. The relevant physical mechanism is that such a mass asymmetry breaks the usual symmetry of the electron and positron thermal distributions, leading to a small chemical potential $\mu_e \sim \delta m$ required to maintain charge neutrality. This, in turn, could influence weak interaction rates, pair densities, and neutrino transport—key ingredients in high-energy astrophysics. However, we find that in all realistic systems, the effect is negligible due to either strong degeneracy, exponential suppression, or dynamic conditions that dominate over thermal equilibrium effects.

\begin{itemize}
    \item \textbf{Core-Collapse Supernovae:} 
    In the inner core of a collapsing massive star, temperatures reach $T \sim 1\text{--}10\ \text{MeV}$, and electrons are highly degenerate with chemical potentials $\mu_e \gg T$, typically in the range $100\ \text{keV} \text{--} 10\ \text{MeV}$ \cite{Janka:2006fh}. Positrons are present in significant numbers, but the system's behavior is governed by the high electron degeneracy and dense nuclear matter. A small mass difference $\delta m \sim \text{keV}$ induces a chemical potential shift of the same order, which is tiny compared to the prevailing $\mu_e$. Consequently, weak interaction rates (e.g., electron and positron capture on nucleons) and neutrino emission remain effectively 
    unchanged \cite{Janka:2006fh, Janka:2012wk}.

    \item \textbf{Neutron Star Mergers:}
    These events also involve high-temperature, high-density matter with significant lepton populations and neutrino fluxes. Like in supernova cores, the matter is partially degenerate and neutrino interactions are important in setting the proton-to-neutron ratio and energy transport. However, the induced chemical potential from a $\text{keV}$-scale mass difference is again far below the thermal and chemical potential scales relevant in such environments \cite{Perego:2019fca}. . Therefore, no appreciable dynamical or observable effect is expected.

    \item \textbf{White Dwarfs and Neutron Star Crusts:}
    These regions have much lower temperatures ($T \sim 10\text{--}100\ \text{keV}$) and are strongly electron-degenerate. Since positron densities are suppressed exponentially as $\sim \exp(-m_{e^+}/T)$, the positron population is negligible, and any mass asymmetry between electrons and positrons has no effect on the equation of state, pressure balance, or reaction rates. The plasma is effectively composed of electrons and ions only, and the positron sector is thermally 
    irrelevant \cite{Shapiro:1983du}.

    \item \textbf{Gamma-Ray Bursts and Pair Plasmas:}
    Some gamma-ray bursts and relativistic jets (e.g., around black holes or in magnetars) may involve pair-dominated, high-temperature plasmas where $T \sim \text{MeV}$ and $e^\pm$ pairs contribute significantly to the thermodynamics \cite{Piran:1993jm, Fermi-LAT:2021zee}. In such settings, a mass difference could, in principle, lead to slight asymmetries in pair creation and annihilation rates, and hence to a tiny net lepton number. However, the systems are rapidly expanding and highly non-equilibrium, so the conditions for the thermal chemical potential to be well-defined are marginal at best. Moreover, the small value of $\delta m / T \sim 10^{-3} \text{--} 10^{-2}$ ensures that any such asymmetry is far subdominant to other uncertainties or dynamics. The resulting effect on observable spectra or plasma dynamics is negligible. 
    
\end{itemize}

\section{Related Work and Complementary Constraints}

Recent work has explored variations in fundamental constants using different astrophysical and cosmological probes. Uniyal et al.~\cite{Uniyal:2025ejp} used a Bayesian neural network on Gaia-DR3 white-dwarf data to constrain variations of the proton-to-electron mass ratio, $\beta$, finding limits of $|\Delta \beta / \beta| = 1.23^{+37.02}_{-35.71} \times 10^{-7}$ when finite-temperature corrections are included in the white dwarf equation of state. For comparison, a $20\%$ increase in the electron mass would correspond to $|\Delta \beta / \beta| = 0.167$, indicating that white dwarf observations provide complementary but less stringent constraints than BBN for the scenarios considered here.

In 2023 Khalife et al. published an updated comparison of Hubble-tension models that include time-varying electron mass. By including either spatial curvature or early dark energy into their models, they were able to reduce the Hubble tension to 1.0-2.9$\sigma$ \cite{Khalife:2023qbu}.

In 2024, Schöneberg et al.~\cite{Schoneberg:2024ynd} investigated the cosmological implications of electron mass variations, finding that small increases in the electron mass can ease the Hubble tension. They provided detailed analysis of how such variations affect CMB observables through modifications to the physics of recombination. The applicability of these constraints to our scenario depends on the temperature at which $m_{e^{\pm}}$ is restored to its SM value. If restoration occurs before recombination, CMB constraints would be significantly weakened, while BBN constraints would remain relevant.

\section{Summary and Conclusions}

We have explored the possibility of temperature-dependent CPT violation through mass differences between electrons and positrons during Big Bang Nucleosynthesis. Our main findings are:

\begin{itemize}
    \item We introduce a framework in which temperature-dependent CPT violation is parametrized by $b_0(T) = \alpha T^2$, where the $T^2$ scaling represents the maximum temperature dependence compatible with both theoretical consistency and phenomenological viability. Higher powers ($T^3$, $T^4$) produce excessive CPT violation at BBN, while lower powers ($T^1$) are either theoretically unnatural or phenomenologically insufficient.
    
    \item We present two toy models demonstrating how $b_0(T) \propto T^2$ scaling can arise from well-motivated field-theoretic structures: a scalar-vector coupling with temperature-driven phase transition (most natural from an EFT perspective) and a PT-symmetric Hamiltonian (alternative non-Hermitian approach).
    
    \item Using the precision BBN code \faGithub \href{https://github.com/vallima/PRyMordial}{\,\texttt{PRyMordial}}, modified to incorporate dynamically-solved chemical potentials, full neutron-proton interconversion rates, and finite-mass corrections to neutrino interactions, we systematically constrain electron-positron mass asymmetries from observed abundances of He-4, D/H, and measurements of $N_{\rm eff}$.
    
    \item BBN observations place stringent constraints on electron-positron mass asymmetries at $T \sim 1$ MeV, corresponding to $\alpha \gtrsim 10^{-6}~\text{GeV}^{-1}$. When considering all three observables (He-4, D/H, $N_{\rm eff}$) simultaneously, we find no combination of electron and positron masses satisfying all constraints at the 1$\sigma$ level. However, pairwise combinations of observables do allow constrained regions of parameter space, indicating sensitivity to the choice of Helium-4 abundance determination and the persistent Lithium Problem. 
    
    \item Current CPT bounds from laboratory experiments apply only at $T \sim 0$ and do not constrain early-universe behavior. The $T^2$ scaling ensures that even large mass differences at BBN ($\sim$ keV scale) correspond to negligibly small effects today ($\lesssim 10^{-35}$ GeV), automatically evading all present-day experimental constraints.
    
    \item The effects of keV-scale electron-positron mass differences in other astrophysical environments (core-collapse supernovae, neutron star mergers, white dwarfs, gamma-ray bursts) are negligible due to either strong degeneracy, exponential suppression, or highly non-equilibrium dynamics.
    
    \item While our primary focus is on BBN constraints, we note that the same framework with much smaller coupling constants ($\alpha \sim 10^{-10}~\text{GeV}^{-1}$) could in principle address baryogenesis through sphaleron conversion at the electroweak scale, though this scenario is physically distinct from and incompatible with the large-$\alpha$ regime constrained by BBN.
\end{itemize}

These results demonstrate that BBN provides the most stringent probe of early-universe CPT violation of this type, constraining a parameter regime inaccessible to laboratory experiments and complementary to other cosmological constraints from the CMB and large-scale structure. The tension between different observables highlights the importance of improved Helium-4 abundance measurements and the need to resolve the Lithium Problem for tighter constraints on new physics during BBN.

\section*{Acknowledgments}

We would like to thank Miguel Escudero Abenza for the illuminating discussions and valuable feedback on our work. GB is supported by the Spanish grants  CIPROM/2021/054 (Generalitat Valenciana) and   PID2023-151418NB-I00 funded by MCIU/AEI/10.13039/501100011033. AKB acknowledges support from the ``Unit of Excellence Maria de Maeztu 2020-2023'' award to the ICC-UB CEX2019-000918-M and grant PID2022-136224NB-C21 funded by MCIN / MINECO / MCOC. AKB and GB are supported by the European Union’s Horizon 2020 research and innovation program under the Marie Sklodowska-Curie grant agreement No 860881-HIDDeN, and Horizon Europe research and innovation programme under the Marie Sk lodowska-Curie Staff Exchange grant agreement No 101086085 – ASYMMETRY.

\end{document}